# Zero-Dimensional Organic-Inorganic Hybrid Material with Ultra-Narrow-Red Emission at Room Temperature

Peiqing Cai[1]*, Song Wang[1], Tianmou Xu[1], Ying Tang[1,2], Xiaolin Yuan[2], Mingjie Wan[1], Qi Ai[1], Junjie Si[1], Xin Yao[1], Yonggang Cao[3], Maxim K. Rabchinskii[4], Pavel N. Brunkov[4] and Zugang Liu[1]*

[1]College of Optical and Electronic Technology, China Jiliang University, Hangzhou, Zhejiang 310018, China.

[2]Ningbo Semiconductor International Corporation, Ningbo, Zhejiang, 315800, China.

[3]College of Metrology and Measurement Engineering, China Jiliang University, Hangzhou, Zhejiang 310018, China.

[4]Ioffe Institute, 26 Politekhnicheskaya, St Petersburg, 194021, Russian Federation.

*Author to whom correspondence should be addressed: Email: pqcai@cjlu.edu.cn and zgliu78@cjlu.edu.cn.

# Abstract

Recently, low-dimensional organic-inorganic hybrid halide compounds have aroused great attention in the optoelectronic field, due to the unique topology and optical properties. Herein, we report an $Mn^{4+}$ doped $[N(CH_3)_4]_2TiF_6$ zero-dimensional organic-inorganic hybrid phosphor, which could not only exhibit very narrow and pure red emission, but also maintain efficient emission intensity at room temperature. The crystal structure, photoluminescence properties and temperature sensing application are discussed. The excellent temperature dependent luminescent properties are attributed to the rigid structure and isolated $MnF_6^{2-}$ octahedra in the total crystal framework. These results will help design suitable materials and devices in both warm white light emitting diodes and optical sensors.

# 1. Introduction

Solid-state white light-emitting diodes (w-LED) have a significant position in the lighting and display markets.[1-4] Currently，the most popular strategy for the w-LED design is the combination of InGaN blue chips and yellow YAG: $Ce^{3+}$ phosphors.[5] However, duo to the intrinsic lack of the red emission, the method above-mentioned could not completely simulate the nature sunlight. Recently, $Mn^{4+}$ doped fluoride phosphors have been proposed as one kind of cost-effective candidates to compensate the red region, which are sensitive to the human eyes.[6-8] The optical absorption and emission properties of $Mn^{4+}$ ions doped inorganic fluoride materials have been known for many years,[9-13] but their wet synthesis method and potential application have not been raised enough attention. A milestone improvement on the application of $Mn^{4+}$ doped fluoride phosphor in w-LED occurred in 2008 by S. Adachi 's group, when they accidentally rediscovered the $Mn^{4+}$ doped $K_2SiF_6$ (KSF) phosphor during the erosion experiment of fluorescent porous silicon.[14, 15] Since then, a various of $Mn^{4+}$ doped inorganic fluoride and oxyfluoride compounds were reported subsequently.[16-22]

Although the $Mn^{4+}$ inorganic fluoride phosphors exhibit narrow pure red emission and broadband absorption in the blue range, the intrinsic hydroscopicity of the $Mn^{4+}$ doped phosphors will trigger the hydrolysis reaction. The manganese fluoride component of the compounds will be hydrolyzed into manganese oxides and hydroxides, which may reduce the emitting intensity and restrict the actual application of the phosphors.[23] Basically, the intense narrow red emission of the $Mn^{4+}$ doped fluoride phosphors is originated from the $MnF_6^{2-}$ octahedral luminescent centers in a specific crystal field environment. Thus, reducing the degradation of the optical active $MnF_6^{2-}$ is the key point to realize the durability in the ambient environment. Several researchers found that the external surface modification was the general method to improve the moisture resistance of the inner delicate $MnF_6^{2-}$ octahedra. For example, Zhou et al. develop a simple surface passivation strategy to treat $KSF:Mn^{4+}$ samples, the surface $Mn^{4+}$ cations are sacrificed by $H_2O_2$, in situ forming a $Mn^{4+}$-rare surface protective layer with low solubility and the $Mn^{4+}$ inside are protected indirectly.[24] Nguyen et al. reported a facile approach for coating $Mn^{4+}$ doped fluoride phosphors with a moisture-resistant alkyl phosphate layer.[25] Huang and coworkers utilized a novel reductive DL-mandelic acid to block the hydrolysis reaction of $MnF_6^{2-}$ clusters.[23]

The zero-dimensional organic-inorganic halide compounds own unique spatial bound structure, in which the inorganic components are isolated by the organic ingredients, thus the luminescent active inorganic components are protected well by the organic ligands, forming a self-modified structure to resist the moisture. Recently, the broadband emissions in zero-dimensional organic-inorganic metal halide perovskites and perovskite-related materials with excellent color tunability and high PLQE have aroused great attention in the w-LED community.[26-28] Although the tunable narrow band emission originated from the excitonic states already have been demonstrated in the 3D, 2D and quasi-2D perovskites,[29-37] narrow band emission in the 0D systems have rarely been reported in the literatures. Recently, T. Jüstel et al reported an organic-inorganic [C(NH$_2$)$_3$]$_2$GeF$_6$: Mn$^{4+}$ phosphor with narrow red emission. However, this material exhibits acute luminescence quenching at room temperature, due to the multi-phonon relaxation induced by the strong phonon modes from the organic guanidinium cations. Therefore, the narrow red emission of the Mn$^{4+}$ could only be observed at cryogenic temperature.[38]

In this work, we reported a Mn$^{4+}$ doped zero-dimensional tetramethylammonium hexafluorotitanium phosphor: [N(CH$_3$)$_4$]$_2$TiF$_6$: Mn$^{4+}$ by the combination of facile solvothermal method and low temperature solid-state method. The as-prepared material not only shows narrow red emission at room temperature, but also exhibits negative luminescent thermal quenching, indicating the practical application potentials in w-LEDs and flexible optoelectronic devices.

## 2. Experiment section

### 2.1 Preparations

A facile wet chemical method was adopted to synthesize [N(CH$_3$)$_4$]$_2$TiF$_6$: Mn$^{4+}$ [tetramethylammonium hexafluorotitanium, (TMA)$_2$TiF$_6$:Mn$^{4+}$] red phosphor. K$_2$MnF$_6$ powders were selected as the Mn$^{4+}$ source and synthesized according to the literatures published elsewhere. [16, 39]. Pure polycrystalline [N(CH$_3$)$_4$]$_2$TiF$_6$ powders were prepared by using a solvothermal method reported in ref [40]. Typically, 0.638 g TiO$_2$, 0.438 g [N(CH$_3$)$_4$]Cl, 20 mL DMF, and 2 mL HF were mixed and stirred for 1 h in a Teflon beaker. The precursor solution was transferred to a hydrothermal reaction vessel and heated to 180 °C for 4 days. After cooling down to room temperature naturally, the crystalline sample was collected by filtration and washed with DMF for three times. The final white product was obtained by heating at a vacuum oven at 120 °C overnight. The $Mn^{4+}$ doped $(TMA)_2TiF_6$ was prepared by a simple solid-state method subsequently. 0.621g $(TMA)_2TiF_6$ and 0.024g $K_2MnF_6$ were completely mixed by grinding in an agate mortar. The mixture was transferred to a home-made Teflon container and heated in the vacuum oven at 100 °C for 3 h.

**2.2 Characterization**

The crystalline information of the powder samples was obtained by means of X-ray diffraction (XRD, D8 Advanced, Bruker Co, Germany) using Cu Kα-radiation (λ = 1.5406 Å), temperature dependent photoluminescence excitation, emission spectra (PLE/PL) and decay curves were measured by utilizing an Edinburgh Instruments FLS1000 spectrometer equipped with a 450W Xe discharge lamp, a 60W μs Xe flash lamp and a closed-cycle Helium cryostat. Fourier transformed infrared (FTIR) spectrum was recorded by a Nicolet 6700 spectrometer with the sample dispersed in a KBr pellet. The Raman spectrum of $Mn^{4+}$ doped sample was measured by the Raman microscope system (Horiba, LabRAM HR).

## 3. Results and discussion

**3.1 Crystallographic determination and structure analysis**

The structure information of $Mn^{4+}$ doped $(TMA)_2TiF_6$ was collected by powder X-ray diffraction and corrected by using Rietveld refinement on the GSAS program.[41] The crystallographic data of the pure $(TMA)_2TiF_6$ reported by Göbel et al was selected as the initial structure model during the structure refinement.[42] Fig. 1(a) exhibits the refined XRD data of 2% $Mn^{4+}$ doped $(TMA)_2TiF_6$ and the detail crystallographic information was listed in Table. 1. As shown in Fig. 1(b), the $Mn^{4+}$ doped $(TMA)_2TiF_6$ is belong to a centrosymmetric trigonal R-3 space group with cell parameters a = b = 7.997 Å, c = 19.915 Å. Isolated $TiF_6^{2-}$ inorganic octahedra were separated by rigid $N(CH_3)_4^+$ organic tetrahedral cations, forming a zero-dimensional molecular framework structure.

Structural vibration spectra studies of the molecular groups in the $(TMA)_2TiF_6$ crystal were performed by FTIR and Raman spectra. The FTIR spectra in Fig. 2 depicted the Ti-F vibrations in the range of 550-620 $cm^{-1}$ with low phonon frequency. Two peaks located at 946 and 1018 $cm^{-1}$ were assigned as the C-N stretching vibrations. The C-H asymmetric vibrations in the $CH_3$ group were observed in the region *ca*. 1452-1512 $cm^{-1}$. In the Raman spectra, the vibration peak observed at 603 $cm^{-1}$ has been assigned as the Ti-F vibration mode, which is in agreement with the result exhibited in the FTIR spectra. Two skeletal modes located at 760 and 958 $cm^{-1}$ has been assigned to the $v_1$ and $v_3$ non-degenerate $A_2$-type modes of the $C_4N$ skeleton, respectively. The C-H $v_{15}$ triply degenerate $F_2$-type vibration was observed at 1486 cm$^{-1}$, which is both infrared and Raman active. These assignments are in agreement with the results in previously reported literatures.[40, 43-45]

### 3.2 Photoluminescence studies

Fig. 3(a) show the normalized PL and PLE spectra of (TMA)$_2$TiF$_6$:Mn$^{4+}$ at room temperature. The PLE spectrum with 630 nm emission exhibits three dominant bands in the range of 200-500 nm. The broad bands peaking at 374 and 470 nm are due to the spin-allowed $^4A_2 \rightarrow {^4T_1}$ and $^4A_2 \rightarrow {^4T_2}$ transitions of Mn$^{4+}$ in octahedral symmetry, while the higher-lying absorption band at 250 nm could be assigned as the charge transfer (CT) band. The room temperature PL in Fig. 3(a) shows typical red sharp lines of Mn$^{4+}$ in a fluoride phosphor, originating from the spin- and parity-forbidden $^2E_g \rightarrow {^4A_{2g}}$ zero phonon line (ZPL) transition and the accompanying Stokes and anti-Stokes vibrational sidebands $v_6$ (630nm, 613nm), $v_4$ (634nm, 609nm) and $v_3$ (6460nm, 598nm), respectively. It has been reported in literatures that there is strong correlation between the ZPL emission intensity and the crystal symmetry of the host materials. S. Adachi has summarized that negligibly small ZPL of Mn$^{4+}$ could be found in the cubic, tetragonal and trigonal structure with high crystal symmetry.[46] As analyzed by the refined structure information in Table 1, the Mn$^{4+}$ ions were confirmed to substitute the TiF$_6^{2-}$ octahedral in a centrosymmetric lattice of trigonal (TMA)$_2$TiF$_6$ system, thus no obvious ZPL was observed at room temperature according to Adachi's experience summary and the practical PL spectra of $(TMA)_2TiF_6:Mn^{4+}$ as shown in Fig. 3(a). However, some vibrational modes from the inorganic octahedral with low energy frequency still could borrow the energy from the $^2E_g \rightarrow {}^4A_{2g}$ ZPL transition, the phonon activated sidebands is therefore located on either side of the ZPL. On the other hand, from the point of view of crystal structure, the luminescent center $MnF_6$ in the zero dimensional $(TMA)_2TiF_6$ does not share any connection with surrounding $N(CH_3)_4^+$ organic tetrahedral by faces, edges or corners, thus the electron-lattice coupling effect, which is usually taken account for the spectral broadening of the phosphors, is trivial in this system. Thus the anti-Stokes and Stokes phonon emission peaks are quite sharp. Fig. 2(b) and (c) exhibit the photograph images of pure and $Mn^{4+}$ doped $(TMA)_2TiF_6$ under illumination of daylight and UV excitation. Efficient red luminescence from the $Mn^{4+}$ doped phosphor could be observed by the UV light illumination. The brown body color of the $Mn^{4+}$ doped phosphor under the daylight illumination also indicates the efficient absorption in the range of UV to blue, corresponding to results revealed in the PLE spectra as shown in Fig. 3(a).

Thermal stability is an important property for phosphors in application. Therefore, the temperature dependent emission spectra and decay curves of the $Mn^{4+}$ doped $(TMA)_2TiF_6$ were investigated. As is shown in Fig.4 (a), the ZPL and Stokes sidebands are dominant in the emission spectra of $Mn^{4+}$ doped $(TMA)_2TiF_6$ at 50 K. When the temperature increases, electrons have more chances of thermal occupancy to populate at higher vibrational state. As a result, the anti-Stokes sidebands were gradually rising on the high energy side of the ZPL. On the contrary with the temperature dependent PL behaviors of $Mn^{4+}$ doped $(TMA)_2TiF_6$, the decay lifetimes of the $Mn^{4+}$ doped phosphor shows significant luminescent quenching. As shown in Fig. 4(b), in the range of the measured temperature, the decay curves are always keep exponential and decay time becomes shorter with temperature increase, indicating the dominated non-radiative relaxation process during the temperature rising experiment. The total average decay time can be calculated as:[47, 48]

$$\tau = \frac{\int_0^\infty tI(t)dt}{\int_0^\infty I(t)dt} \quad (1)$$

where $I$ is the intensity at time $t$, $t$ is the time, and $\tau$ is the average decay time. The calculated values of the lifetimes at various temperatures have been labeled in Fig. 4(b). Because the temperature dependent kinetic processes of the $Mn^{4+}$ $^2E_g \rightarrow {}^4A_{2g}$ emission are very complicated in the full temperature range from cryogenic temperature to room temperature, the excited dynamics in the $Mn^{4+}$ doped fluoride phosphors should comprehensively consider several factors such as activation energy, spin-orbit interaction and vibration modes, etc.[46, 49, 50] A valid and simple non-radiative transition theory model with a few fitting parameters is difficult to be built to describe the temperature dependent behaviors of the lifetimes of the $Mn^{4+}$ doped phosphors.

Fig. 4 (c) shows the integrated emission intensities of anti-Stokes sidebands, Stokes sidebands and total $Mn^{4+}$ emission as functions of temperature in the temperature range of 50 to 300 K. The intensity of all the components of the emission increased with increasing temperature from 50 to 300 K. As ascribed above, the anti-Stokes vibrational state and Stokes vibrational state could be simply regarded as two coupled vibronic levels, which can borrow the energy from ZPL. The temperature dependent fluorescence intensity ratio (FIR) from the thermal coupling levels (TCL) of active $Mn^{4+}$ ions can be modified as: [51, 52]

$$FIR = Aexp(-\Delta E/kT) \qquad (2)$$

where $A$ is constant, $\Delta E$ is the energy difference between two coupled levels, $k$ is the Boltzmann constant, and $T$ is the temperature. Fig. 4 (c) illustrates the intensity ratio $I_{anti-S}$ / $I_S$ of $Mn^{4+}$ emission as a function of temperature in the range of 50 to 300 K. The sharp increase in $I_{anti-S}$ / $I_S$ with increasing temperature indicates that the $Mn^{4+}$ doped organic-inorganic phosphor has a potential application in temperature sensors. We delete the baseless description here

The temperature dependent PL spectroscopy data represented here is quite different with the that of $Mn^{4+}$ doped $[C(NH_2)_3]_2GeF_6$ (CGF) phosphors reported by T. Jüstel et al.[38] The $Mn^{4+}$ red luminescence in CGF host is quenched at room temperature and they assumed that the low thermal quenching temperature is caused by the multi-phonon relaxation induced by the high phonon frequencies of the organic component in the lattice (3400 $cm^{-1}$, N-H stretching vibration). According to ref [53], the energy of C-H stretching vibration in the $[N(CH_3)_4]_2TiF_6$ host is about 3000 $cm^{-1}$, comparable with the energy of the N-H stretching vibration in CGF. If this assumption is correct, the $Mn^{4+}$ doped $[N(CH_3)_4]_2TiF_6$ should exhibit similar luminescent properties as $Mn^{4+}$ doped CGF phosphor. In this work, the $Mn^{4+}$ doped $[N(CH_3)_4]_2TiF_6$ exhibits no luminescence quenching when it is heated from low temperature to the room temperature as shown in Fig. 4(c). To investigate the reasons behind this difference, we synthesized the $Mn^{4+}$ doped $[C(NH_2)_3]_2TiF_6$ phosphor for comparison, which has an isostructure with the CGF crystal. We found that, as same as $Mn^{4+}$ doped CGF, the as-synthesized $[C(NH_2)_3]_2TiF_6$:$Mn^{4+}$ powders also show no red emission at room temperature under UV excitation. Ok et al. [40] investigated the structure differences between the centrosymmetric $[N(CH_3)_4]_2TiF_6$ and nocentrosymmetric $[C(NH_2)_3]_2TiF_6$, and they found that both materials have the zero-dimensional crystal structure consisting $TiF_6$ that are separated by organic cations. However, the stereo $N(CH_3)_4^+$ tetrahedra cations could help to build a rigid structure framework of the $[N(CH_3)_4]_2TiF_6$ matrix, whereas the nitrogen atoms in planar $C(NH_2)_3^+$ have significant hydrogen bonding with the fluorine atoms in $TiF_6$, giving $[C(NH_2)_3]_2TiF_6$ a 'soft' pseudo two-dimensional topology. As is well known, the $3d^3$ $Mn^{4+}$ ion is belong to high field ions in a particular electronic configuration and the PL properties of $Mn^{4+}$ are strongly influenced by the crystal environment, covalency and electron cloud expansion (*nephelauxetic* effect), etc.[54, 55] Therefore, we suggest that the ordered rigid structure with strong ionicity in $[N(CH_3)_4]_2TiF_6$ could effectively maintain the transition metal coordination field and suppress phonon scattering, thereby promoting the radiative transition at elevated temperature. On the contrary, the pseudo two-dimensional structure in $[C(NH_2)_3]_2TiF_6$ or $[C(NH_2)_3]_2GeF_6$ cannot effectively block the movement of high-energy phonons, and the strong hydrogen bond interaction may also influence the high-order perturbation terms and coupling representation, reducing the transition probability of the radiative matrix element, thus the red emission of $Mn^{4+}$ is fully quenched at room temperature. Furthermore, we also synthesized the $Mn^{4+}$ doped $(4,4'\text{-bpyH}_2)TiF_6$ powders, in which the isolated $TiF_6$ octahedra are localized in the hydrogen bond framework created by the bipyridinium cations.[56] No red emission was observed at room temperature in this material, which is in agreement with the hypothesis of hydrogen bond effect mentioned above.

## 4. Conclusions

A new red $Mn^{4+}$ doped $[N(CH_3)_4]_2TiF_6$ phosphor has successfully synthesized by a mild solvothermal method for the first time. The as-prepared $Mn^{4+}$ doped phosphor presented sharp red line emission of $Mn^{4+}$ at around 630 nm under excitation in the near UV to blue region at room temperature. The temperature dependent PL and decay of $Mn^{4+}$ were performed to investigate the origin of the efficient red emission at room temperature. Spectroscopy analyses demonstrated that enhanced phonon-assistant emission and elevated non-radiative transition are two competitive processes during the temperature rising process, which are responsible for the higher PL intensity and shorter decay time at elevated temperature. Comparative study of structure and bonding interaction on $[N(CH_3)_4]_2TiF_6$ and $[C(NH_2)_3]_2TiF_6$ also indicated that the $Mn^{4+}$ doped organic-inorganic hybrid materials with rigid structure and strong ionic bond were necessary for efficient red emission. Our findings highlight the importance of judiciously choosing organic cations in the $Mn^{4+}$ doped materials for the emission at room temperature, which has guiding significance for the design of narrow red organic-inorganic hybrid luminescent materials..

# CRediT author statement

**Peiqing Cai:** Conceptualization, Writing-Original draft preparation. Supervison. **Song Wang:** Software, Resources. **Tianmou Xu:** Software, Resources. **Ying Tang:** Data Curation. **Xiaolin Yuan:** Data Curation. **Mingjie Wan:** Resources. **Qi Ai:** Visualization. **Junjie Si:** Investigation. **Xin Yao:** Investigation. **Yonggang Cao:** Investigation. **Maxim K. Rabchinskii:** Writing-Review & Editing. **Pavel N. Brunkov:** Writing-Review & Editing. **Zugang Liu**: Writing-Review & Editing. Supervison.

# Conflicts of interest

There are no conflicts to declare.

# Acknowledgements

This work was supported by the National Natural Science Foundation of China (NSFC) (11904345, 61905230, 11904344), the Natural Science Foundation of Zhejiang Province (LQ19F040004, LQ20A040007), the Scientific Research Foundation of China Jiliang University (000629) and Zhejiang Province International Scientific Cooperation Project (Ningbo 2018D10007).

Figure

Fig. 1 (a) Rietveld refinement XRD patterns of the (TMA)$_2$TiF$_6$: 2% Mn$^{4+}$. (b) Schematic illustration of (TMA)$_2$TiF$_6$ and coordination environment of TiF$_6^{2-}$ octahedral.

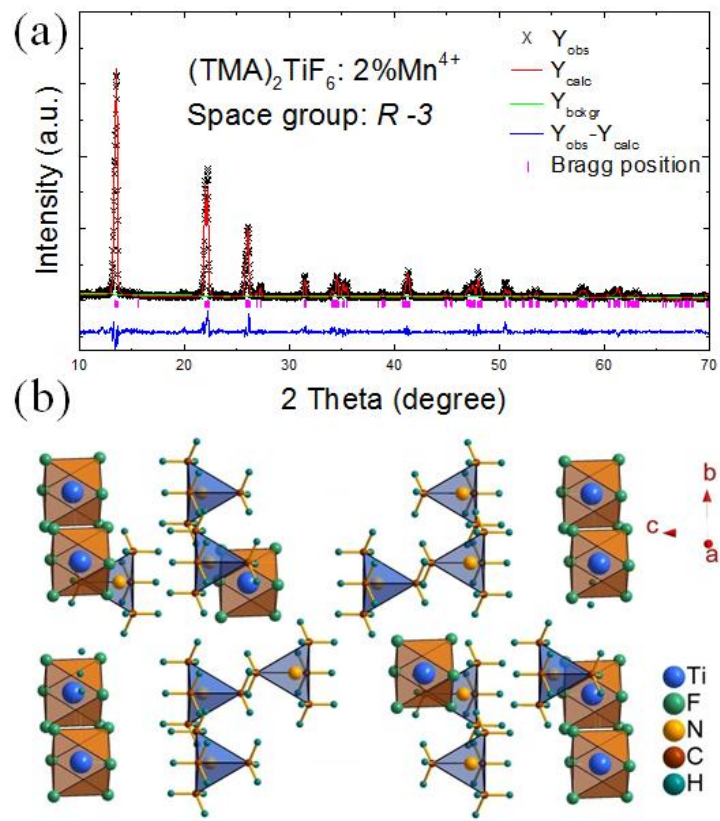

Fig. 2 Room temperature FTIR spectrum and Raman spectrum of pure (TMA)$_2$TiF$_6$.

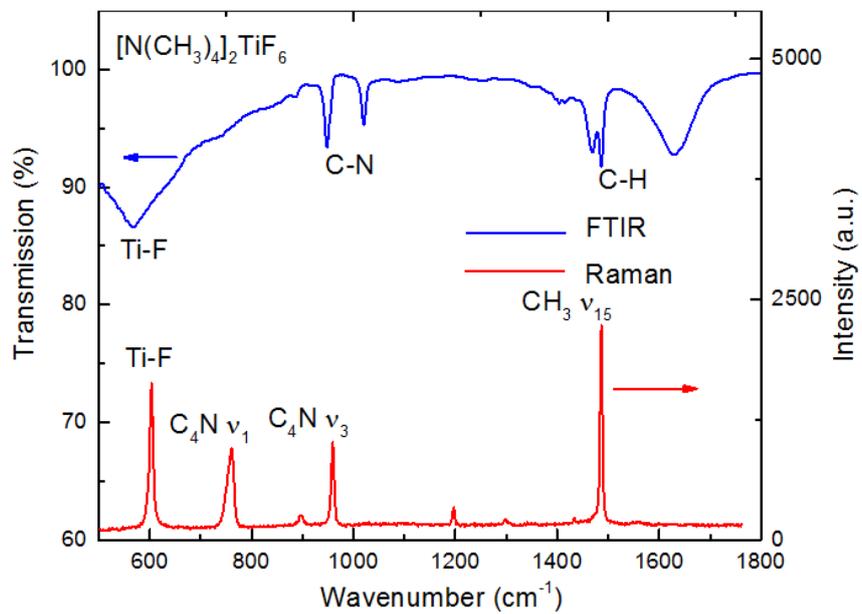

Fig. 3 (a) PLE and PL spectra of the $(TMA)_2TiF_6:Mn^{4+}$ at room temperature. (b) and (c) photograph images of pure and $Mn^{4+}$ doped $(TMA)_2TiF_6$ under illumination of daylight and UV excitation, respectively.

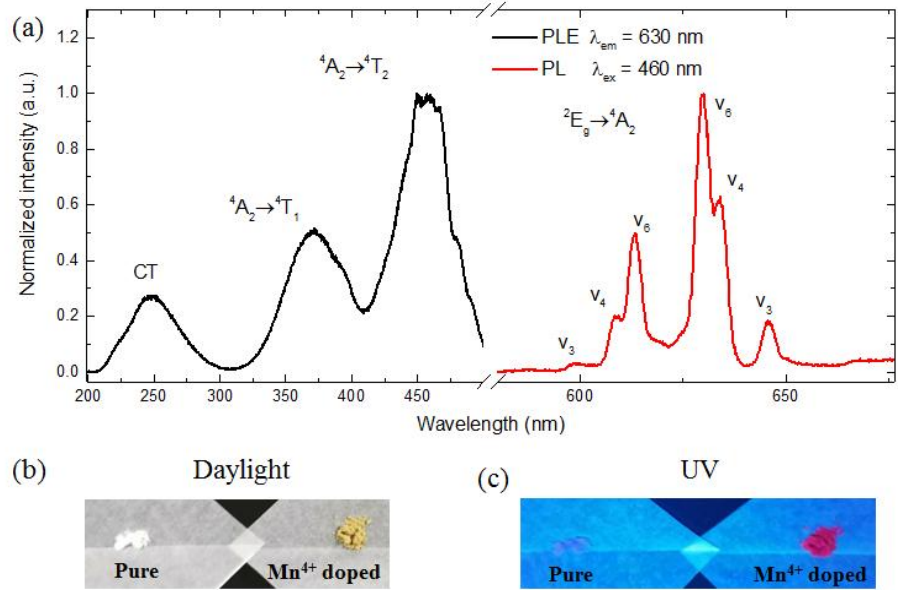

Fig. 4 (a) Temperature dependent PL spectra of $Mn^{4+}$ doped $(TMA)_2TiF_6$ as a function of temperature under 465 nm excitation in the range from 50 K to 300 K. (b) Temperature dependent decay curves of $Mn^{4+}$ $(TMA)_2TiF_6$. (c) Calculated integrated emission intensity of total $^2E_g \rightarrow {}^4A_{2g}$ transition ($I_{total}$), anti-Stokes ($I_{a-S}$), and Stokes ($I_S$) phonon sidebands, respectively. (c) Relationship between $I_{a-S}$ and $I_S$ as a function of temperature.

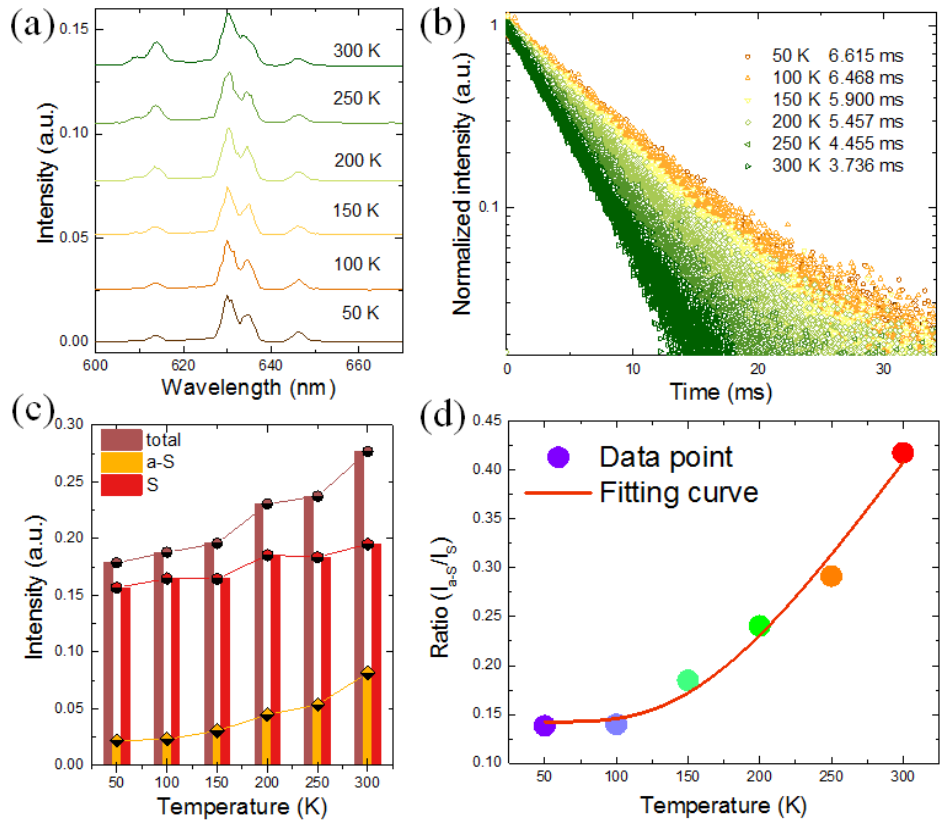

Table

Table 1. Refined atomic coordinate parameters data of $(TMA)_2TiF_6:2\%Mn^{4+}$ at room temperature.

| space group: R-3: a = 7.998 Å. b = 7.998 Å, c = 19.915 Å, α = 90°, γ = 120°. | | | | | | |
|---|---|---|---|---|---|---|
| atom | site | x | y | z | occupacy | $U_{iso}$ (Å$^2$) |
| Ti | 3a | 0.0000 | 0.0000 | 0.0000 | 1 | 0.0172 |
| F | 18f | 0.1490 | 0.2134 | 0.0536 | 1 | 0.0318 |
| N | 6c | 0.0000 | 0.0000 | 0.2485 | 1 | 0.0092 |
| C1 | 6c | 0.0000 | 0.0000 | 0.3233 | 1 | 0.0247 |
| H1 | 18f | 0.1300 | 0.0529 | 0.3393 | 1 | 0.0025 |
| C2 | 18f | 0.1317 | 0.2000 | 0.2234 | 1 | 0.0209 |
| H2 | 18f | 0.1321 | 0.1998 | 0.1751 | 1 | 0.0900 |
| H3 | 18f | 0.2602 | 0.2444 | 0.2396 | 1 | 0.0038 |
| H4 | 18f | 0.0874 | 0.2847 | 0.2392 | 1 | 0.0493 |
| $R_{wp}$=19.27 %, $R_p$=14.6%, $\chi^2$=7.54, vol. = 1273.923 Å$^3$. | | | | | | |

Structure Directing Properties of Polar Metal Oxyfluoride [MOxF6–x]2–(x= 1, 2) Building Units, Inorganic Chemistry, 54 (2015) 1712-1719.